\input{aipcheck}

\documentclass[
    ,final            % use final for the camera ready runs
  ]
  {aipproc}

\layoutstyle{6x9}

\begin{document}

\title{Present-Day Star Formation: Protostellar Outflows 
and Clustered Star Formation
}


\author{Fumitaka Nakamura}{
  address={National Astronomical Observatory of Japan; 
Mitaka, Tokyo 181-8588, Japan; fumitaka.nakamura@nao.ac.jp}
}

\author{Zhi-Yun Li}{
  address={Department of Astronomy, University of Virginia, 
P.O. Box 400325, Charlottesville, VA 22904, USA; zl4h@virginia.edu}
}

%\author{<author3>}{
%  address={<common address for author2 and author3>}
%  ,altaddress={<author1 address>} % additional visiting address
%}

\begin{abstract}
Stars form predominantly in clusters inside dense clumps of 
turbulent, magnetized molecular clouds. 
The typical size and mass of the cluster-forming clumps are 
$\sim 1$ pc and $\sim 10^2 - $ 10$^3$ M$_\odot$, respectively.
Here, we discuss some recent progress on theoretical and
observational studies of clustered star formation 
in such parsec-scale clumps with emphasis on the role of protostellar 
outflow feedback.
Recent simulations indicate that protostellar outflow feedback 
can maintain supersonic turbulence in a cluster-forming 
clump, and  the clump can keep a virial equilibrium long
after the initial turbulence has decayed away.
In the clumps, star formation proceeds relatively slowly; it continues for 
at least several global free-fall times of the parent dense 
clump ($t_{\rm ff}\sim $ a few $\times 10^5$ yr). 
The most massive star in the clump is formed at the bottom of the clump
gravitational potential well at later times through the filamentary 
mass accretion streams that are broken up by the outflows 
from low-mass cluster members.
Observations of molecular outflows in nearby cluster-forming clumps
appear to support the outflow-regulated cluster formation model.
\end{abstract}

\maketitle

%%%%%%%%%%%%%%%%%%%%%%%%%%%%%%%%%%%%%%%%%%%%
%% MAINMATTER
%%%%%%%%%%%%%%%%%%%%%%%%%%%%%%%%%%%%%%%%%%%%

\section{Introduction}

Most stars form in clusters (Lada \& Lada 2003). 
Therefore, understanding the formation process of star clusters is 
a key step towards a full understanding of how stars form.
Recent observations have revealed that star clusters form 
in turbulent, magnetized, parsec-scale dense clumps of molecular clouds
(Ridge et al. 2003).
These clumps contain masses of $10^2-10^3 M_\odot$,
fragmenting into an assembly of cores 
that collapse to produce stars. 
In cluster-forming clumps, stellar feedback such as protostellar outflows, 
stellar winds, and radiation rapidly start to shape 
the surroundings. 
Because of the short separations between forming stars 
and cores, these feedback mechanisms are expected to control 
subsequent star formation.

Among the stellar feedback processes, protostellar outflow feedback
has been considered to be one of the important mechanisms
that control the structure and dynamical properties of cluster-forming
clumps (e.g., Matzner \& McKee 2000) because the outflows from a
group of young stars interact with a substantial volume of their
parent clump by sweeping up the gas into shells. 
Furthermore, in nearby cluster-forming clumps which are forming 
no massive stars that would emit strong UV radiation, the protostellar 
outflow feedback is considered to be the dominant feedback mechanism.
Recent 
numerical simulations of cluster formation have demonstrated
that the protostellar outflows largely regulate the structure
formation and star formation in a dense cluster-forming clump.
Li \& Nakamura (2006) showed that the supersonic turbulence 
in dense clumps can be maintained by the momentum injection 
from the protostellar outflows (see also Carroll et al. 2009). 
The moderately-strong magnetic fields are also important
to impede the rapid global gravitational collapse.
In this case, the global inflow and outflow are expected to coexist,
interacting with themselves. As a result, the cluster-forming clumps 
as a whole can keep quasi-equilibrium states for a relatively long time.
Furthermore, Nakamura \& Li (2007) showed that the global star 
formation efficiency tends to be reduced by the momentum injection 
from the protostellar outflows, although local star formation 
can often be triggered by the dynamical compression due to 
the protostellar outflows. 

In this contribution, we discuss several important characteristics of 
this outflow-regulated cluster formation model and 
compare the theoretical model with observations of nearby
cluster-forming regions.

\section{Outflow-Regulated Cluster Formation}

Most, perhaps all, stars go through a phase of vigorous outflow 
during formation. Li \& Nakamura (2006) examined, 
through 3D MHD simulation, the effects of protostellar outflows 
on cluster formation. They found that the initial turbulence 
in the cluster-forming region is quickly replaced by motions 
generated by outflows. The protostellar outflow-driven turbulence 
(``protostellar turbulence'' for short) can keep the region close 
to a virial equilibrium long after the initial turbulence has 
decayed away. Therefore, there exist two types of turbulence 
in star-forming clouds: a primordial (or ``interstellar'') turbulence 
and a protostellar turbulence, with the former transformed into
the latter mostly in embedded clusters such as NGC 1333. 
Collimated outflows are more efficient in driving turbulence 
than spherical outflows that carry the same amounts of momentum. 
This is because collimated outflows can propagate farther
away from their sources, effectively increasing the turbulence 
driving length; turbulence driven on a larger scale
decays more slowly. Gravity plays an important role in 
shaping the turbulence, generating infall motions that
balance the outward motions driven by outflows.
Since the majority of stars are thought to form in clusters,
an implication is that the stellar initial mass function is 
determined to a large extent by the stars themselves, through
outflows that individually limit the mass accretion onto 
forming stars and collectively shape the environments
(density structure and velocity field) in which most cluster members
form. Thus, massive cluster-forming clumps supported by protostellar 
turbulence gradually evolve toward a highly centrally condensed 
``pivotal'' state, culminating in rapid formation of massive stars 
in the densest part through accretion (see below for massive 
star formation).
Here, we call this cluster formation scenario ``outflow-regulated
cluster formation''.

\subsection{Star Formation Rate in Cluster Formation}

Using 3D MHD numerical simulations of cluster formation,
Li \& Nakamura (2006) and Nakamura \& Li (2007) demonstrated that 
the protostellar outflow-driven turbulence can keep 
a pc-scale, cluster-forming clump close to a virial equilibrium long
after the initial turbulence has decayed away.
Here, we derive an analytic formula of star formation rate
in a cluster-forming clump that keeps its virial equilibrium 
by the protostellar outflow feedback. 

Numerical simulations of protostellar turbulence indicate that 
the dissipation rate of the turbulence momentum, 
$dP_{\rm turb}/dt$, balances 
the momentum injection rate by the protostellar outflow feedback, 
$dP_{\rm out}/dt$, so that the clump can be kept close to a virial equilibrium. 
Adopting the virial equilibrium condition, Nakamura \& Li (2011)
derived the predicted star formation rate per free-fall time
(SFR$_{\rm ff}$) toward observed cluster-forming
clumps as follows.
\begin{equation}
{\rm SFR}_{\rm ff} 
\simeq 0.0125 \alpha \left(\frac{f_{\rm B}}{0.5}\right) 
\left(\frac{f_{\rm w}}{0.5}\right)^{-1}
\left(\frac{V_{\rm w}}{10^2 \ {\rm km \ s^{-1}}}\right)^{-1}
\left(\frac{R_{\rm cl}}{1\ {\rm pc}}\right)^{1/2}
\left(\frac{\Sigma_{\rm cl}}
{5 \times 10^{21} {\rm cm^{-2}}}\right)^{1/2} \ .
\end{equation}

\begin{figure}
\includegraphics[height=.3\textheight]{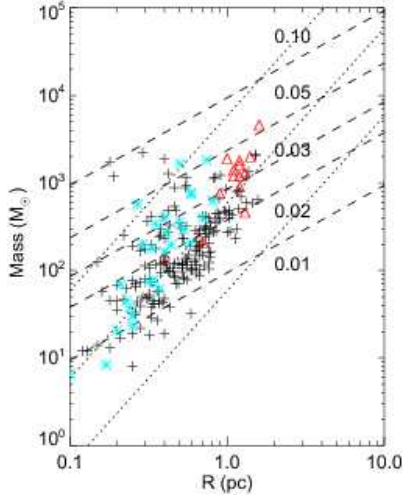}
\caption{Contours of SFR$_{\rm ff}$ predicted by
the outflow-regulated cluster formation model on the mass-radius
diagram. The dashed contours are labeled by values of SFR$_{\rm ff}$.
The dotted  contours indicate the constant column density
at $10^{21}$ cm$^{-2}$ ({\it bottom line}), $10^{22}$ ({\it intermediate
 line}), 
and $10^{23}$ cm$^{-2}$ ({\it upper line}).
The crosses and triangles indicate the cluster-forming clumps 
identified by Ridge et al. (2003) and Higuchi et al. (2009), respectively.
The asterisks indicate the dense clumps in the IRDCs
identified by Rathborne et al. (2006).
For most of the dense clumps, the predicted SFR$_{\rm ff}$ ranges 
from 1 \% to 5 \%. See Nakamura \& Li (2011) in more detail.
}  
\end{figure}

Figure 1 shows the dependence of SFR$_{\rm ff}$ 
on the mass and radius.
The outflow-regulated cluster formation model suggests that the star formation
rate per free-fall time ranges from 1 \% to 5 \% 
for the observed cluster-forming clumps in the solar neighborhood
when the protostellar outflow feedback maintains the supersonic turbulence 
in the clumps, indicating that it takes about (2 $-$ 10) $t_{\rm ff}$
for the star formation efficiency to reach about (10 $-$ 20) \%.

\subsection{Massive Star Formation}

\begin{figure}
  \includegraphics[height=.3\textheight]{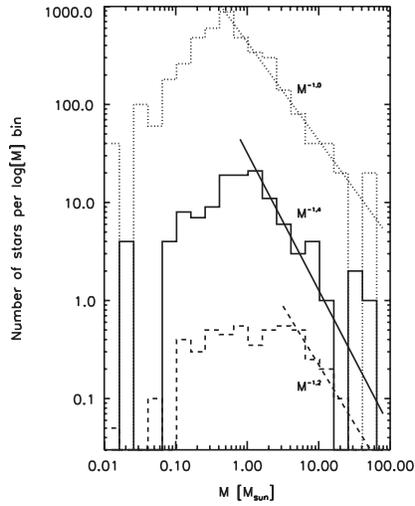}
\caption{Power-law fits to the high mass end of the stellar mass 
distributions of the three models: HD (without outflow and magnetic field,
dashed lines), MHD (without outflow, solid), and WIND (with outflow and
 magnetic field, dotted). The top (bottom) curve is raised 
(lowered) by a factor of 20 for clarity.
See Li et al. (2010) in more detail.
}  
\end{figure}

In the outflow-regulated cluster formation, massive stars 
are predicted to form in later phases of cluster formation
at the bottom of the global gravitational potential, or
at the central densest part.
Wang et al. (2010) investigated massive star formation in turbulent, 
magnetized, parsec-scale clumps of molecular clouds including
protostellar outflow feedback using 3D numerical simulations 
of effective resolution $2048^3$. The calculations are carried out 
using a block structured adaptive mesh refinement code, ENZO, that 
solves the ideal magnetohydrodynamic equations including 
self-gravity and implements accreting sink particles. 
They found that, in the absence of regulation by magnetic fields 
and outflow feedback, massive stars form readily in a turbulent, moderately
condensed clump of about 1600$M_\odot$ (containing about 100 
initial Jeans masses), along with a cluster of hundreds of lower
mass stars. The massive stars are fed at high rates by 
(1) transient dense filaments produced by large-scale turbulent
compression at early times and (2) by the clump-wide global 
collapse resulting from turbulence decay at late times
(see also Smith et al. 2009).
In both cases, the bulk of the massive star's mass is supplied 
from outside a 0.1 pc-sized core that surrounds the
star. In the simulation, the massive star is clump-fed 
rather than core-fed. The need for large-scale feeding makes
the massive star formation prone to regulation by outflow feedback, 
which directly opposes the feeding processes.
The outflows reduce the mass accretion rates onto the massive stars 
by breaking up the dense filaments that feed the
massive star formation at early times, and by collectively slowing 
down the global collapse that fuels the massive
star formation at late times. The latter is aided by a moderate 
magnetic field of strength in the observed range
(corresponding to a dimensionless clump mass-to-flux ratio $\sim$ a
few); 
the field allows the outflow momenta
to be deposited more efficiently inside the clump. 
Thus, the massive star formation in the simulated
turbulent, magnetized, parsec-scale clump is outflow-regulated 
and clump-fed. An important implication is that the
formation of low-mass stars in a dense clump can affect the formation 
of massive stars in the same clump, through
their outflow feedback on the clump dynamics.

In addition, contrary to the common expectation, a magnetic field of the
strength in the observed range decreases, rather than increases, 
the characteristic stellar mass. It (1) reduces the number of 
intermediate-mass stars that are formed through direct turbulent 
compression, because sub-regions of the clump with masses 
comparable to those of stars are typically magnetically 
subcritical and cannot be compressed directly into collapse, 
and (2) increases the number of low-mass stars that are produced 
from the fragmentation of dense filaments. The filaments result 
from mass accumulation along the field lines. 
In order to become magnetically supercritical and fragment, 
the filament must accumulate a large enough column density 
(proportional to the field strength), which yields a high volume density 
(and thus a small thermal Jeans mass) that is conducive to forming
low-mass stars. 
The characteristic stellar mass is reduced further by outflow feedback.
The conclusion is that both magnetic fields and outflow feedback are 
important in shaping the stellar initial mass function
(see Figure 2).

\subsection{Physical Properties of Dense Cores in Cluster Formation}

\begin{figure}
\includegraphics[height=.21\textheight]{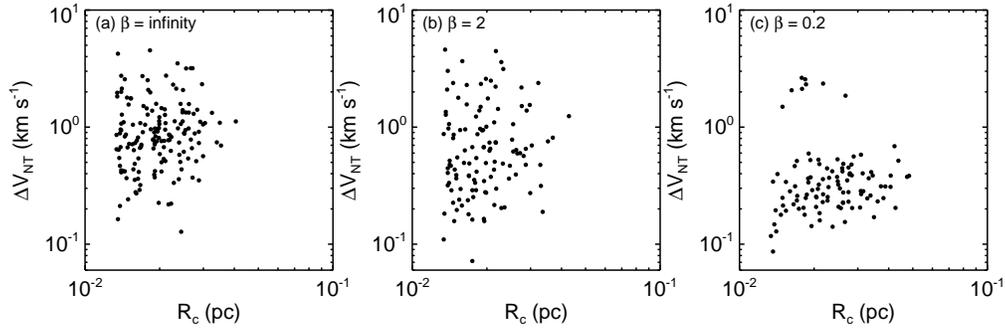}
\caption{Nonthermal 3D velocity dispersions 
as a function of core radii for three models with different 
magnetic field strengths: (a) model N1 (no magnetized),
(b) model W1 (weakly magnetized), and 
(c) model S1 (strongly magnetized). 
The sound speed is $c_s$ = 0.266 km s$^{-1}$, 
corresponding to $T = 20$ K.
See Nakamura \& Li (2011) in more detail.}
\end{figure}

In the outflow-regulated cluster formation model, 
the physical properties of dense cores in cluster-forming clumps 
are also different from those in quiescent star-forming 
regions like Taurus.
Nakamura \& Li (2011) have performed a set of 3D MHD simulations
of cluster formation taking into account the effects of
protostellar outflow feedback and identified dense cores by applying
a clumpfind algorithm to the simulated 3D density data cubes. 
The main results are summarized as follows.
(1) The velocity dispersions of dense cores show little correlation 
with core radius, irrespective of the strength of the magnetic field 
and outflow feedback (see Figure 3). In the absence of a magnetic field, 
the majority of the cores have
supersonic velocity dispersions, whereas in the presence of
a moderately strong magnetic field, the cores tend to be
subsonic or at most transonic.
(2) Most of the cores are out of virial equilibrium,
with the external pressure due to ambient turbulence
dominating the self-gravity. The core formation and evolution
is largely controlled by the dynamical compression
due to outflow-driven turbulence. Such a situation is in contrast
to the strongly magnetized (magnetically subcritical)
case, where the self-gravity plays a more important role in
the core dynamics, particularly for massive cores (Nakamura \& Li 2008).
(3) Even an initially weak magnetic field can retard star formation
significantly, because the field is amplified by supersonic
turbulence to an equipartition strength. In such an
initially weak field, the distorted field component dominates
the uniform one. In contrast, for a moderately strong
field, the uniform component remains dominant. Such a
difference in the magnetic structure can be observed in
simulated polarization maps of dust thermal emission. Recent
polarization measurements show that the field lines in
nearby cluster-forming clumps are spatially well ordered,
indicative of a moderately strong, dynamically important,
field (see Figure 5b, e.g., Sugitani et al. 2010, 2011).

The characteristics of dense cores formed in dense clumps 
with moderately strong magnetic fields are in good agreement 
with observations of $\rho$ Oph, the nearest cluster-forming clump
(Maruta et al. 2010), where subsonic or transonic internal motions 
are observed in dense cores, most of which appear 
gravitationally-unbound or pressure-confined
(see Figure 4).

\section{Comparison with Observations}

\begin{figure}
  \includegraphics[height=.23\textheight]{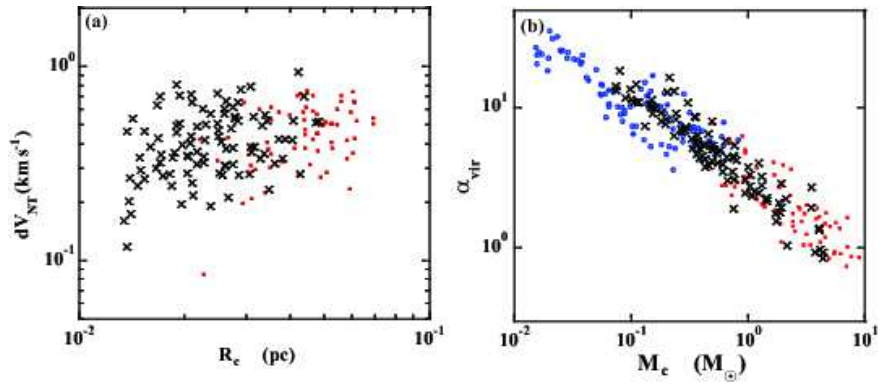}
\caption{The relationship between the 1D FWHM nonthermal 
velocity width and core radius for the dense cores 
in the $\rho$ Ophiuchi Main Cloud and our cores identified 
from the moderately strong magnetic field model (model S1). 
The red dots and black crosses denote the $\rho$ Oph cores 
and our cores of model S1, respectively. For the $\rho$ Oph cores, 
we used the results of Maruta et al. (2010) who identified 
the cores using H$^{13}$CO$^+$ ($J=1-0$) data.
(b) The virial-parameter-core-mass relation. The red dots and 
black crosses are the same as those of panel (a). 
The blue open circles denote the cores of NGC 1333 
identified using N$_2$H$^+$ ($J=1-0$) emission (Walsh et al. 2007). 
See Nakamura \& Li (2011) in more detail.}
\end{figure}

Recently, several extensive molecular outflow surveys towards cluster
forming regions have been done using the molecular outflow tracers, CO
lines.  Here, we summarize the survey results toward 5 nearby 
cluster-forming clumps. Since all 5 regions contain no massive stars 
emitting strong UV, the outflow feedback is expected to be the 
leading stellar feedback mechanism. 
An example of these outflow surveys is shown in Figure 5, 
where blueshifted and redshifted  CO ($J=3-2$) integrated intensity 
contours are indicated towards the nearest cluster-forming, infrared 
dark cloud, Serpens South ($d\sim 400 $ pc).
Several molecular outflow lobes are crowded in the central dense region
where about 100 YSOs are located and forming a protocluster.
The overlapping outflow structure is a common characteristic of active 
cluster forming clumps, suggesting that the outflows 
influence the structure formation in the clumps significantly.

Table 1 summarizes some physical properties of the outflows 
for the 5 regions.
From the observational data, we can now address the two important 
issues of the cluster formation: turbulent generation and clump destruction.
If we compare between the turbulent dissipation rate and outflow energy
injection rate, we can verify whether the outflow feedback can 
maintain the supersonic turbulence in the clumps.
As for the clump destruction, we can compare the global 
gravitational force and the force exerted by the outflows. 
The outflow force must be comparable to the clump gravitational 
force if the outflow feedback plays an important role in the clump destruction. 

\begin{figure}
  \includegraphics[height=.24\textheight]{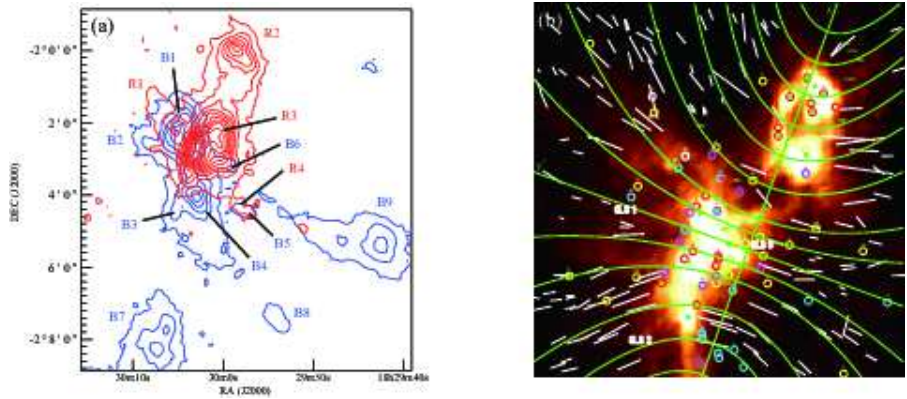}
\caption{(a) Molecular outflow lobes identified from CO ($J=3-2$) 
emission toward the Serpens South protocluster. 
The blue contours represent blueshifted
$^{12}$CO gas and red contours represent redshifted $^{12}$CO gas. 
The blue and red contour levels go up in 6 K km s$^{-1}$ steps, 
starting from 3 K km s$^{-1}$. The integration ranges
are -9.75 to 3.75 km s$^{-1}$ for blueshifted gas and 
11.25 to 29.25 km s$^{-1}$ for redshifted gas.
See Nakamura et al. (2011b) in more detail.
(b) H-band polarization vectors (white lines) toward the Serpens Main cloud, 
and their best-fit magnetic field (green thick curved/straight lines 
and thin curved lines), superposed on SCUBA 850 $\mu$m continuum 
image. The horizontal axis (green thin straight line), which is 
perpendicular to the parabolic magnetic field axis 
(green thick straight line), is also shown. 
YSOs identified by the Spitzer Space Telescope are marked by 
circles (Class 0/I; red, Flat spectrum; magenta, Class II; yellow, 
Transition disk; blue, Class III; white) with ID numbers 
(Tables 3 and 4 of Winston et al. 2007). Submillimeter continuum
peaks are shown by crosses with names and their coordinates come 
from the Spitzer photometry of Winston et al. (2007), 
except SMM2 and SMM11 (Davis et al. 1999).
}
\end{figure}

According to Table 1, for all the clumps, the outflow energy injection
rate is comparable to or larger than the turbulence dissipation rate. 
So, we conclude that the outflows can maintain supersonic turbulence 
in the cluster-forming clumps. On the other hand, 
how the outflow force impacts the global clump dynamics appears to
depend on the clump mass.
For the clumps with masses larger than about 400 $M_\odot$,
the clump gravitational force tends to be significantly larger than 
the outflow force, whereas for the clumps with masses smaller 
than about 400 $M_\odot$, the outflow force is comparable to
the clump gravitational force.
It may be difficult to destroy the whole clumps 
only by the outflow feedback for the clumps with masses larger than 
400 $M_\odot$. 
In contrast, the current observed outflow activity
can destroy the small clumps or at least change the clump dynamics 
significantly.
Since the typical YSO ages are around a few Myr, 
the cluster formation may last for several or 
more free-fall times. 
That supports the slow cluster formation scenario or 
the outflow-regulated cluster formation scenario. 
However, the importance of the outflow feedback in cluster formation
may depend on the clump mass and therefore further detailed 
investigation will be needed to uncover the role of protostellar outflow
feedback in cluster formation.

%%%%%%%%%%%%%%%%%%%%%%%%%%%%%%%%%%%%%%%%%%%%
%% Sample figure:
%%
%% The option [height=...] scales the picture to the given height,
%% without it it would be printed at its nominal size
%%%%%%%%%%%%%%%%%%%%%%%%%%%%%%%%%%%%%%%%%%%%

%%%%%%%%%%%%%%%%%%%%%%%%%%%%%%%%%%%%%%%%%%%%
%% SAMPLE TABLE
%%
%% Shows the use of \tablehead and \tablenote
%% macros
%%%%%%%%%%%%%%%%%%%%%%%%%%%%%%%%%%%%%%%%%%%%

\begin{table}
\begin{tabular}{lrrrrrrrr}
\hline \tablehead{1}{l}{b}{Name \\}
  & \tablehead{1}{l}{b}{Mass \\ ($M_\odot$)}
  & \tablehead{1}{l}{b}{$\dot{E}_{\rm turb}$ \\ ($L_\odot$)}
  & \tablehead{1}{l}{b}{$\dot{E}_{\rm out}$  \\ ($L_\odot$)}
  & \tablehead{1}{l}{b}{$\dot{E}_{\rm out}/\dot{E}_{\rm turb}$  \\}
  & \tablehead{1}{l}{b}{${F}_{\rm grav}$ \\ ($M_\odot \ \rm km/s$) }
  & \tablehead{1}{l}{b}{${F}_{\rm out}$ \\ ($M_\odot \ \rm km/s$)}
  & \tablehead{1}{l}{b}{${F}_{\rm out}/{F}_{\rm grav}$ \\}
  & \tablehead{1}{l}{b}{Ref.\tablenote{1: Nakamura et al. (2012), 2:
     Graves et al. (2010), 3: Nakamura et al. (2011b), 4: Nakamura et
     al. (2011a), 5: Arce et al. (2010)} \\}   \\
\hline
L1641N & 200 & 0.1 & 2  & 20 & 6 & 10 & 1.7 & 1\\
Serpens Main & 400 & 0.2 & 1 & 10 & 2 & 5 & 2.5 & 2\\
Serpens South & 500 & 0.2 & 1 & 5 & 40 &8 & 0.2 & 3 \\
rho Oph & 1000 & 0.1 & 0.2 & 2  & 13& 1 & 0.08 & 4 \\
NGC 1333 &2000 & 0.5 & 0.7 & 1.4  & 20 & 6 & 0.3 & 5 \\
\hline
\end{tabular}
\caption{Observations of nearby parsec-scale cluster-forming clumps}
\label{tab:a}
\end{table}

%%%%%%%%%%%%%%%%%%%%%%%%%%%%%%%%%%%%%%%%%%%%%%%%
%% BACKMATTER
%%%%%%%%%%%%%%%%%%%%%%%%%%%%%%%%%%%%%%%%%%%%%%%%

\bibliographystyle{aipproc}


%%\bibliographystyle{aipprocl} % if natbib is missing
%
%
%%%%%%%%%%%%%%%%%%%%%%%%%%%%%%%%%%%%%%%%%%%%
%%% You probably want to use your own bibtex database here
%%%%%%%%%%%%%%%%%%%%%%%%%%%%%%%%%%%%%%%%%%%%
%\bibliography{sample}

\begin{thebibliography}{9}
\bibitem{arce10}
Arce, H. G., Borkin, M. A., Goodman, A. A., Pineda, J. E., \& Halle, M. W.
2010, ApJ, 715, 1170
\bibitem{carroll09}
Carroll, J. J., Frank, A., Blackman, E. G., et al. 2009, ApJ, 695, 1376
\bibitem{davis99}
Davis, C. J., et al. 1999, MNRAS, 309, 141
\bibitem{graves10}
Graves, S. F., et al. 2010, MNRAS, 409, 1412
\bibitem{lada03}
Lada, C. J., \& Lada, E. A. 2003, ARA\&A, 41, 57
\bibitem{li06}
Li, Z.-Y., \& Nakamura, F. 2006, ApJ, 640, L187 
\bibitem{li10}
Li, Z.-Y., Wang, P., Abel, T., \& Nakamura, F. 2010, ApJ, 720, L26
\bibitem{higuchi09}
Higuchi, A. E., Kurono, Y., Saito, M., \& Kawabe, R. 2009, ApJ, 705, 468
\bibitem{maruta10}
Maruta, H., Nakamura, F., Nishi, R., Ikeda, N., \& Kitamura, Y. 
2010, ApJ, 714, 680
\bibitem{matzner00}
Matzner, C. D., \& McKee, C. F. 2000, ApJ, 545, 364
\bibitem{matzner07}
Matzner, C. D. 2007, ApJ, 659, 1394
\bibitem{nakamura07}
Nakamura, F., \& Li, Z.-Y. 2007, ApJ, 662, 395
\bibitem{nakamura08}
Nakamura, F., \& Li, Z.-Y. 2008, ApJ, 687, 354
\bibitem{nakamura11}
Nakamura, F., \& Li, Z.-Y. 2011, ApJ, 740, 36
\bibitem{nakamura11a}
Nakamura, F., Kamada, Y., Kamazaki, T.,  et al. 2011a, ApJ, 726, 46
\bibitem{nakamura11b}
Nakamura, F., Sugitani, K., Shimajiri, Y., et al. 2011b, ApJ, 737, 56
\bibitem{Nakamura2012}
Nakamura, F., Miura, T., Kitamura, Y. et al., 2012, ApJ, 746, 25
\bibitem{rathborne06}
Rathborne, J. M., Jackson, J. M., \& Simon, R. 2006, A\&A, 641, 389
\bibitem{ridge03}
Ridge, N. A., Wilson, T. L., Megeath, S. T., Allen, L. E., \& 
Myers, P. C. 2003, AJ, 126, 286
\bibitem{smith09}
Smith, R. J., Longmore, S., Bonnell, I. 2009, MNRAS, 400, 1775
\bibitem{sugitani10}
Sugitani, K., Nakamura, F., Tamura, M., et al. 2010, ApJ, 716, 299
\bibitem{sugitani11}
Sugitani, K., Nakamura, F., Watanabe, M., et al. 2011, ApJ, 734, 63
\bibitem{walsh07}
Walsh, A. J., Myers, P. C., Di Francesco, J., et al. 2007, ApJ, 655, 958
\bibitem{wang10}
Wang, P., Li, Z.-Y., Abel, T., \& Nakamura, F. 2010, ApJ, 709, 27
\bibitem{winston07}
Winston, E., et al. 2007, ApJ, 669, 493
\end{thebibliography}
%
%%%%%%%%%%%%%%%%%%%%%%%%%%%%%%%%%%%%%%%%%%%%
%%% Just a reminder that you may have to run bibtex
%%% All of it up to \end{document} can be removed
%%% if you don't like the warning.
%  \typeout{******************************************}
%  \typeout{** Please run "bibtex \jobname" to optain}
%  \typeout{** the bibliography and then re-run LaTeX}
%  \typeout{** twice to fix the references!}
%  \typeout{******************************************}

\end{document}